\newcommand{\teff}{T$_{\mathrm{eff}}$}
\newcommand{\Teff}{T$_{\mathrm{eff}}$}
\newcommand{\logg}{log~$g$}
\newcommand{\feh}{[Fe/H]}
\newcommand{\oh}{[O/H]}
\newcommand{\mgh}{[Mg/H]}

\newcommand{\Mgrad}{$\Delta$[Fe/H]/$\Delta$R}
\newcommand{\mygrad}{-0.0678}
\newcommand{\Vgrad}{$\Delta$[Fe/H]/$\Delta$Z}

\newcommand{\acronym}[1]{{\small{#1}}}

\newcommand{\project}[1]{\textsl{#1}}
\newcommand{\gaia}{\project{Gaia}}

\newcommand{\apogee}{\project{\acronym{APOGEE}}}
\newcommand{\aspcap}{\project{\acronym{ASPCAP}}}
\newcommand{\astronn}{\texttt{astroNN}}
\newcommand{\lamost}{\project{\acronym{LAMOST}}}

\newcommand{\degree}{$^{\circ}$}
\newcommand{\hii}{\hbox{{\rm H}\kern 0.1em{\sc ii}{\rm }}}

\documentclass[twocolumn]{aastex631}

\begin{document}

\title{[X/Fe] Marks the Spot: Mapping Chemical Azimuthal Variations in the Galactic Disk with APOGEE}
\shorttitle{Chemical azimuthal variations in \apogee\ DR17}

\author[0000-0002-3855-3060]{Zoe Hackshaw}
\affiliation{Department of Astronomy, The University of Texas at Austin,
2515 Speedway Boulevard 
Austin, TX 78712, USA}
\email{zoehackshaw@utexas.edu}

\author[0000-0002-1423-2174]{Keith Hawkins}
\affiliation{Department of Astronomy, The University of Texas at Austin,
2515 Speedway Boulevard 
Austin, TX 78712, USA}

\author[0000-0001-5522-5029]{Carrie Filion}
\affiliation{Department of Physics \& Astronomy, The Johns Hopkins University, 
Baltimore, MD 21218, USA}
\affiliation{Center for Computational Astrophysics, Flatiron Institute, 162 5th Ave., 
New York, NY 10010, USA}

\author[0000-0003-1856-2151]{Danny Horta}
\affiliation{Center for Computational Astrophysics, Flatiron Institute, 162 5th Ave., 
New York, NY 10010, USA}

\author[0000-0003-3922-7336]{Chervin F. P. Laporte}
\affiliation{Institut de Ci\`encies del Cosmos (ICCUB), Universitat de Barcelona (IEEC-UB), 
Mart\'i i Franqu\`es 1, E-08028 Barcelona, Spain}

\author[0000-0002-5840-0424]{Christopher Carr}
\affiliation{Department of Astronomy, Columbia University, 550 West 120th Street, 
New York, NY 10027, USA}

\author[0000-0003-0872-7098]{Adrian M. Price-Whelan}
\affiliation{Center for Computational Astrophysics, Flatiron Institute, 162 5th Ave., 
New York, NY 10010, USA}

\begin{abstract}

Chemical cartography of the Galactic disk provides insights to its structure and assembly history over cosmic time. In this work, we use chemical cartography to explore chemical gradients and azimuthal substructure in the Milky Way disk with giant stars from APOGEE DR17. We confirm the existence of a radial metallicity gradient in the disk of \Mgrad\ $\sim \mygrad\ \pm 0.0004$ dex/kpc and a vertical metallicity gradient of \Vgrad\ $\sim -0.164 \pm 0.001$. We find azimuthal variations ($\pm0.1$ dex) on top of the radial metallicity gradient that have been previously established with other surveys. The APOGEE giants show strong correlations with stellar age and the intensity of azimuthal variations in \feh; young populations and intermediate-aged populations both show significant deviations from the radial metallicity gradient. Beyond iron, we show that other elements (e.g., Mg, O) display azimuthal variations at the $\pm0.05$ dex-level across the Galactic disk. We illustrate that moving into the orbit-space could help constrain the mechanisms producing these azimuthal metallicity variations in the future. These results suggest that dynamical processes play an important role in the formation of azimuthal metallicity variations.


\end{abstract}



\section{Introduction} \label{sec:Introduction}

The field of Galactic archaeology exists to answer long-standing questions about the processes that drive Galactic formation and evolution \citep{Eggen1962,Searle1978}. We can use the Milky Way and its resolved Galactic components as a laboratory to answer questions about Galactic evolution and characterize the hierarchical formation \citep{Davis1985} of the Milky Way.

The current investigation of the Galactic processes that drive the Milky Way's evolution has exploded within the last few decades due to wide-field missions aimed at mapping the stellar content of the Milky Way such as APOGEE \citep{Majewski2017} and Gaia \citep{GaiaCC22}. Employing stars as the key witnesses to Galactic evolution, we are able to take stellar information (stellar parameters, kinematics, chemical abundances, etc.) and apply them with spatial information to create information-dense maps of the Milky Way in a process known as chemical cartography \citep{Hayden2015}. Measuring the distribution of elements throughout the Milky Way disk can inform us about global and secular processes that dominate over space and time \citep[e.g.,][]{Hawkins2015}. 

To postulate which Galactic phenomena are the most influential, observations of the Galactic disk (both thick and thin) are needed to identify which signatures and patterns prevail. One of the most prominent trends in the Milky Way is the existence of the negative radial and vertical metallicity gradients \citep[an incomprehensive list includes:][]{Mayor1976,Andrievsky2002,Magrini2009,Luck2011,Bergemann2014,Xiang2015,Yan2019,Hawkins2022}. These gradients could provide supporting evidence for certain formation theories of the Milky Way, such as the `inside-out' formation theory \citep{Larson1976}. This theory suggests that the negative radial (\Mgrad) metallicity gradient in the Milky Way could be caused by the inner Galaxy forming first and fast, leading to a higher metallicity concentration towards the Galactic center. As time proceeds, the outer Galaxy starts to form causing it to be metal-poor compared to its centralized counterpart \citep{Matteucci1989,Chiappini1997}. 

While the metallicity gradients are some of the most prominent features observed in the Galaxy, there are more subtle chemical characteristics that appear to be washed out by these strong trends. Chemical azimuthal substructure in the Milky Way has been previously identified using a variety of different tracers such as \hii\ regions \citep{Balser2011,Balser2015} and Cepheids \citep{Lemasle2008,Pedicelli2009}. Recently, using large-scale stellar surveys, azimuthal variations in \feh\ throughout the disk are quantifiable on the level of $\sim 0.1$ dex \citep{Hawkins2022,Poggio2022,imig2023}. 

One motivator for looking for angle-dependent chemical variations in our Galaxy is the confirmed existence of azimuthal variations in galaxies beyond the Milky Way. \citet{Ho2017} found \oh\ azimuthal substructure in NGC 1365, proposing that the spiral arms in this galaxy plays a large role in the dispersion of metals throughout the galaxy. \citet{Hwang2019} show clear azimuthal metallicity variations in SDSS IV MaNGA galaxies in Figure 13.  \citet{Kreckel2019} observed subtle azimuthal variations in 4 out of 8 in their sample of nearby galaxies from the PHANGS–MUSE survey, with ranging associations between the metallicity variations and the spiral arms. This range in correlations with the spiral arms calls into question if spiral arms are the most adequate explanation for the causes of azimuthal chemical variations.

Azimuthal metallicity variations can be generated by different mechanisms in different stellar populations. In older populations, dynamical processes are predicted to be the most important. In younger stars, dynamical processes are thought to have less of an effect on their orbits due to these stars interacting less with features in the Galaxy that can alter their orbits. If azimuthal metallicity variations were present in these younger populations, they would trace any variations in the gas from which they were born \citep{Spitoni2019}.

There are two distinct dynamical mechanisms that spiral arms can generate to induce azimuthal metallicity variations. The first process is radial migration/churning \citep{sellwood2002} which, in this context, is the morphing of stellar orbits that alters the angular momentum of an orbit without the addition of excess energy. The second process is kinematic heating/blurring which would induce changes (heat) in the stellar orbital parameters without increasing the angular momentum. An example of this is the spiral arms dynamically evoking changes in the motions of stars along the leading and trailing edges of the spiral arms \citep{Grand2012,Grand2016}.

In simulations, \citet{Khoperskov2018} found that azimuthal variations in metallicity may arise from the dynamics of stellar disks alone, without the need of radial migration to reshape the stellar population. This is due to dynamically cooler populations in the disk showing a larger contribution to spiral arms than dynamically hotter populations, leading to azimuthal variations in metallicity. \citet{Khoperskov2021} claimed that in various phase-space coordinates, stars in an angular momentum overdensity caused by the spiral arms also exhibit a mean metallicity that is systematically higher than stars not in an angular momentum overdensity. 

Using \gaia\ DR3 stars, \citet{Poggio2022} found that the azimuthal variations present in the disk correlate with where the spiral arms are predicted to be using different samples of bright stars (using effective temperature as a proxy for age) within $\sim4$ kpc of the Sun. When considering older stellar populations, \citet{Dimatteo2013} provides context illustrating that the strength of the azimuthal variations is highly dependent on the strength of the perturbations.


\citet{Debattista2024} quantified azimuthal metallicity variations ($\delta$[Fe/H]) in a Milky Way-like galaxy simulation and found variations that are comprable with the magnitude of the azimuthal variations seen in the Galaxy. Their azimuthal variations in metallicity were coincident with the spiral density waves and present in both young and old populations of stars. When looking at the pattern speeds of the $\delta$[Fe/H] variations, they found that these pattern speeds matched those of the spirals, pointing to the spiral arms as the root cause of these variations. 


There are a handful of explanations, aside from spiral arms, for these azimuthal variations that have been introduced in the literature. One possibility for these angle-dependent trends could be secular processes, such as influence and migration due to the Galactic bar \citep{Dimatteo2013}. Specifically, it has been speculated that the presence of azimuthal variations in an old stellar population is a probe of bar activity. Using $N$-body simulations, \citet{filion23} found that the bar induces azimuth-dependent trends in stellar radii, which then coincides with angle-dependent variations in metallicity using younger to intermediate-age stars ($\sim1 - 4$ Gyr).

A recent study done suggested that the azimuthal variations could arise from interactions with a satellite galaxy, such as Sagittarius \citep{Carr2022}. Torques from the gravitational interaction between an external satellite and the disk of the host galaxy can cause radial migration of the stellar population where inward-migrating stars would be more metal-poor on average compared to in-situ populations and outward-migrating populations would be more metal-rich than the in-situ outer Galactic population (with an assumed negative radial metallicity gradient). This migration response will be induced with respect to the location of the perturber, which then stimulate the azimuthal variations between inward and outward migrating populations. With the authors' simulations, they posited that the influence of Sagittarius will be the most prominent in the outer disk, a conclusion also drawn from \citet{Laporte2018b}. Observationally, \citet{Hwang2019} found gas-phase azimuthal metallicity variations in close or interacting galactic pairs, providing evidence for merger-induced azimuthal metallicity variations.

Most likely, there will be a contribution from all of the aforementioned dynamical processes that could cause azimuthal variations. Empirical observations are necessary to try to disentangle the processes that are responsible for the observed azimuthal metallicity variations in the Galactic disk. While the azimuthal variations of \feh\ in the Galactic disk have been characterized by different tracer populations such as \apogee\ red giants \citep{Eilers2022} and LAMOST OBAF-type stars as well as \gaia\ \citep{Hawkins2022,Poggio2022}, little work has been done when looking beyond iron towards other elements, investigating the effect that stellar population age has on the intensity of azimuthal variations, and examining the dynamical perspective.

 \begin{figure}
	 \includegraphics[width=1\columnwidth]{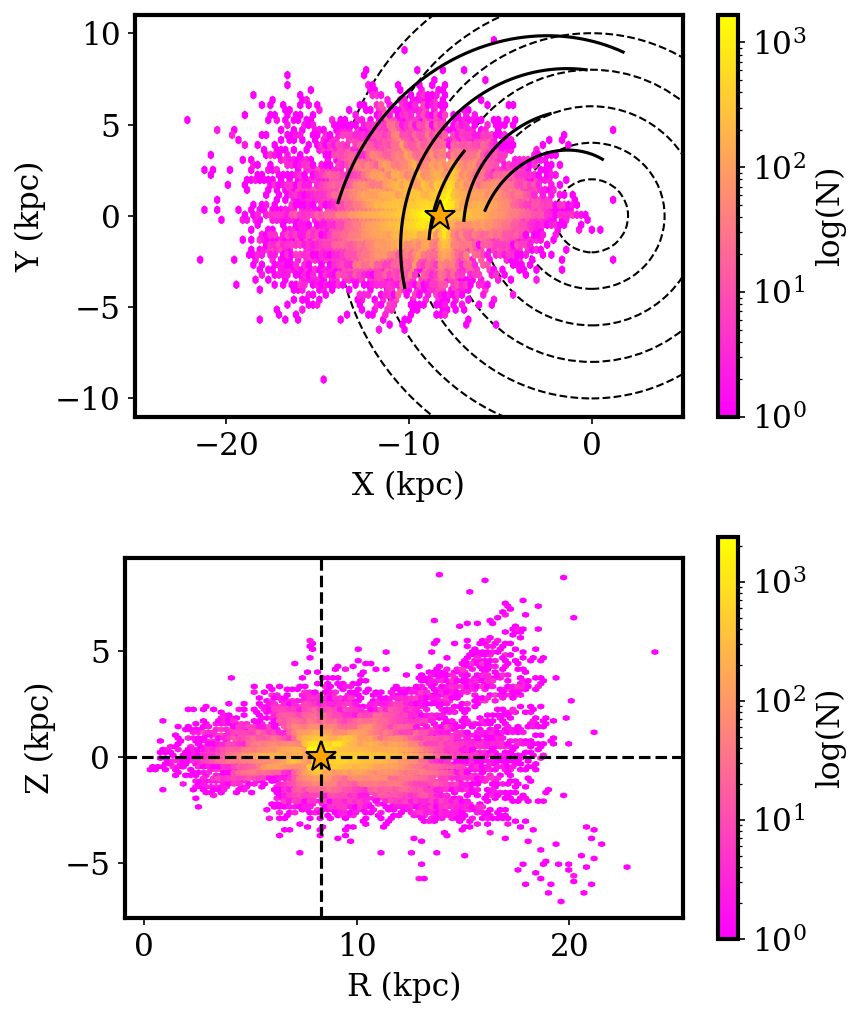}
	\caption{The face-on (top) and edge-on (bottom) distribution of our thin disk sample of 202,510 stars in which the the hexagonal bins are colored by the logarithmic number of stars. The black contours in the top panel are the spiral arms of the Milky Way determined by \citet{Reid2019}. The orange star in both panels represents the Sun. } 
	\label{fig:xy vs rz}
\end{figure}

Throughout this work, we aim to confirm the azimuthal variations in the Galactic disk and investigate any correlation with the spiral arms of the Galaxy. We additionally split our sample by age to quantify the effect that stellar age has on the intensity of the azimuthal variations, examine azimuthal variations in elements beyond Fe, and probe the dynamical origins of these azimuthal metallicity variations. In Section \ref{sec:data}, we delineate the dataset adopted for this project. In Section \ref{sec:Methods}, we explain the methods taken to achieve our goals. In Section \ref{sec:radial_grad}  we present our radial and vertical \feh\ gradients, and highlight the azimuthal deviations from the radial gradient in Section \ref{subsec:azvars}. We quantify azimuthal variations in multiple other elements in Section \ref{subsec:otherelems}. We separate our sample into distinct age bins and characterize how stellar populations of different ages have disparities in their azimuthal variations in Section \ref{subsec:ages}. In Section \ref{subsec:kinematics}, we link the stellar chemistry to kinematics. We summarize our results in Section \ref{sec:summary}.


\section{Data}
\label{sec:data}

The initial sample of stars came from the seventeenth data release \citep[DR17,][]{apogeedr17} of the Apache Point Observatory Galactic Evolution Experiment (\apogee) \citep{apogee}. \apogee\ is a large scale near-infrared (15140\AA\ \textless\ $\lambda$ \textless\ 16960\AA) stellar spectroscopic survey. The survey spans both hemispheres, consisting of a spectrograph \citep{Wilson2019} on the 2.5-meter Sloan Foundation Telescope  at the Apache Point Observatory in New Mexico \citep{Gunn2006}, USA as well as the 2.5-meter Irénée du Pont Telescope at the Las Campanas Observatory in Chile \citep{Bowen73}. Detailed explanation of the data processing and reduction can be found in \citet{Nidever2015}.

We made use of the \astronn\ value-added catalog \citep{leung2019}\footnote{The \astronn\ package is available here: \url{https://github.com/henrysky/astroNN}} of abundances, distances, and ages for \apogee\ DR17 sources. \astronn\ is an open source python package developed for the neural network trained on the \apogee\ data and is designed to be a general package for deep learning in astronomy. The \apogee-\astronn\ catalog contains results from applying \astronn\ neural nets on \apogee\ DR17 spectra to infer stellar parameters, abundances trained with \aspcap\ DR17 \citep[Holtzman et al., in prep.;][]{GarciaPerez2016,Shetrone2015,Smith2021}, distances retrained with \gaia\ eDR3 \citep{Gaia_edr3} from \citet{LeungBovy2019}, and ages trained with APOKASC-2 \citep{Mackereth2019} in combination with low-metallicity asteroseismic ages \citep{Montalban2021}. 

The neural network from \citet{leung2019} mimics a `standard' spectroscopic analysis by using the full wavelength range to deduce stellar parameters and specific sections of the spectrum to derive individual elemental abundances. \astronn\ takes into account incomplete and noisy training data while applying dropout variational inference to find uncertainties on the measurements. This catalog contains stellar parameters (\teff, \logg, and \feh) as well as 18 individual element abundances with precisions at the $\sim0.03$ dex level, agreeing quite well with the traditional \aspcap\ pipeline and producing a smaller scatter with tighter uncertainties. 

For the distances in this work, we adopted the weighted distance in \apogee-\astronn. This parameter is a weighted combination of the \astronn\ distance \citep[spectro-photometric calibrated distances from][]{LeungBovy2019} and Gaia parallax \citep{Gaiasummary2022}. The Galactocentric X, Y, and Z coordinates were found as transformations from the Galactocentric cylindrical radius, azimuth, and vertical height given by \astronn. The Galactocentric positions and velocities were computed assuming the Sun is located at 8.125 kpc from the Galactic center \citep{Gravity2018}, 20.8 pc above the Galactic midplane \citep{BennettBovy2019} and has radial, rotational, and vertical velocities of -11.1, 242, and 7.25 km/s, respectively \citep{Schonrich2010,Bovy2012}.

To explore the effect of stellar age on the chemical azimuthal variations, we sorted our sample into 3 distinct groups with the ages in \apogee-\astronn. Due to the tendency to under-predict old ages, \citet{leung2019} and \citet{Mackereth2019} used a non-parametric Locally Weighted Scatterplot Smoothing (LOWESS) to correct the ages. While this inconsistency mainly applies to stars older than $8+$ Gyr, we followed the recommendation and applied the LOWESS corrected ages for the analysis in this work. 


To explore the intersection of chemistry and dynamics, we aim to quantify how metallicity excess ($\delta$[Fe/H]) changes with respect to the dynamical parameters radial action (J$_r$), angular momentum (L$_z$) which is proportional to the azimuthal action (J$_{\phi}$), vertical action (J$_z$), orbital eccentricity (e), maximum height above the Galactic plane (Z$_{\text{max}}$), and total energy. These parameters were found by integrating the stellar orbits with the \texttt{Gala} code designed primarily for Galactic dynamics evaluations \citep{gala}. The mass-model of the Milky Way used for the \texttt{gala.dynamics.orbit} function was the \texttt{MilkyWayPotential2022} that has been fit to the rotation curve in \citet{Eilers2019} and incorporates the phase-space spiral in the solar neighborhood set by \citet{DarraghFord2023}. 

\begin{figure}
	 \includegraphics[width=1\columnwidth]{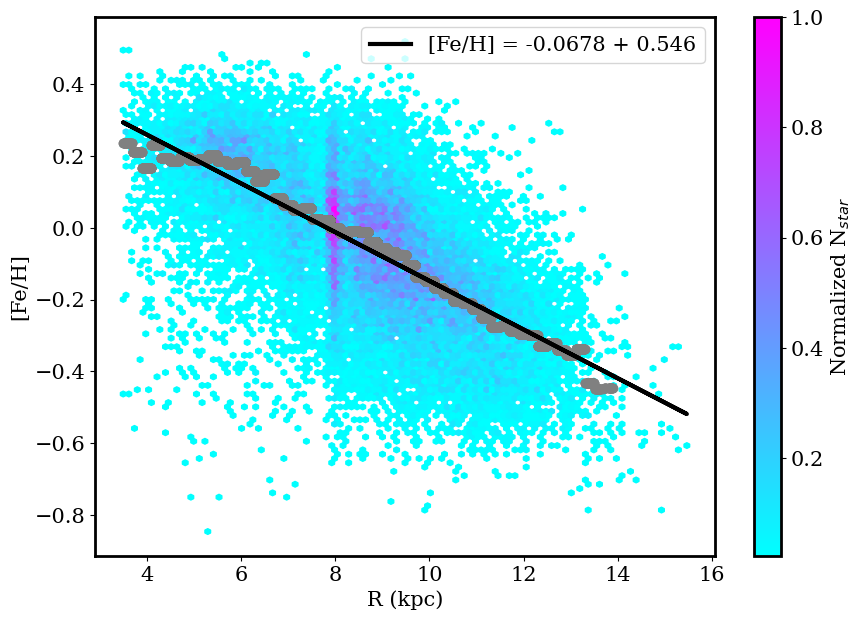}
	\caption{ The \feh\ abundances with respect to Galactocentric radius of our planar thin disk sample. The artifact at R $\sim8$ kpc is an observational effect of the over-representation of stars in the solar neighborhood. We fit a linear model to the colored data points (column-normalized planar thin disk sample) which is represented by the black line. The grey points are the running medians of [Fe/H] in 0.2 kpc bins. From this, we obtain a metallicity gradient of \Mgrad\ $\sim \mygrad\ \pm 0.0004$ dex/kpc and a y-intercept of 0.546 dex for the stars in our planar thin disk sample. } 
	\label{fig:feh vs r}
\end{figure}

With the \apogee-\astronn\ catalog, we employed the following cuts to obtain a set of stars that is well-sampled and a reliable representation of the population we aim to characterize:

\begin{enumerate}
\setlength\itemsep{1em}
\item The Galactocentric distances are an integral part of this work to characterize the metallicity gradient in the Galactic disk. To ensure we have precise positions we place an error cut on the distance measurements obtained from the \astronn\ dataset, selecting stars with errors $ < 30\% $.

\item To minimize systematic effects and dwarf contamination in our sample, we removed stars outside of the effective temperature range $3500 <$ \Teff\ $< 5000$ K as well as any star with \logg\ $> 3.6$ dex. This selection criterion establishes that we are sampling the true red giant section of the Color-Magnitude Diagram (CMD).

\item We removed any stars with an error larger than 0.08 dex on \feh, \oh, or \mgh\ due to these elements being the most relevant in our chemical cartography. The limit of 0.08 dex was chosen based on previous studies that have found the dispersion of line-to-line abundances in \apogee\ is $\sim0.08$ dex for \feh\ \citep{Chen2015,Hawkins2016b} We placed no selection cut on any of the other elements to maintain a balance between the robustness and size of our sample. For completeness we include all of the elements in Section \ref{fig:allgrads} but we refrain from drawing any conclusions about Galactic evolution from the elements with no error cuts (elements other than \feh, \oh, and \mgh).

\item Chemical composition and evolution varies differently among the thin disk, thick disk, and halo. We select only the kinematic thin disk stars based on the processes outlined in \citet{Ramirez2013} (full derivation shown in Appendix \ref{appendix:thindisk}). In this work, the authors outlined a probabilistic approach to assign membership probabilities for stars to thin disk, thick disk, or halo populations based on their Galactic space velocities. We required all thin disk stars to have a TD/D probability $>80\%$, providing us with a final sample of 202,510 stars, shown in Figure \ref{fig:xy vs rz}.
\end{enumerate}

With this full thin disk sample, we are able to characterize the vertical metallicity gradient as a function of Galactocentric radius as well as the radial metallicity gradient as a function of height above the plane. To investigate the chemical gradients and azimuthal variations in the plane of the disk, we further sub-sample the 202,510 stars in the following section.

\subsection{Planar Thin Disk Sample}

\begin{figure}	\includegraphics[width=1\columnwidth]{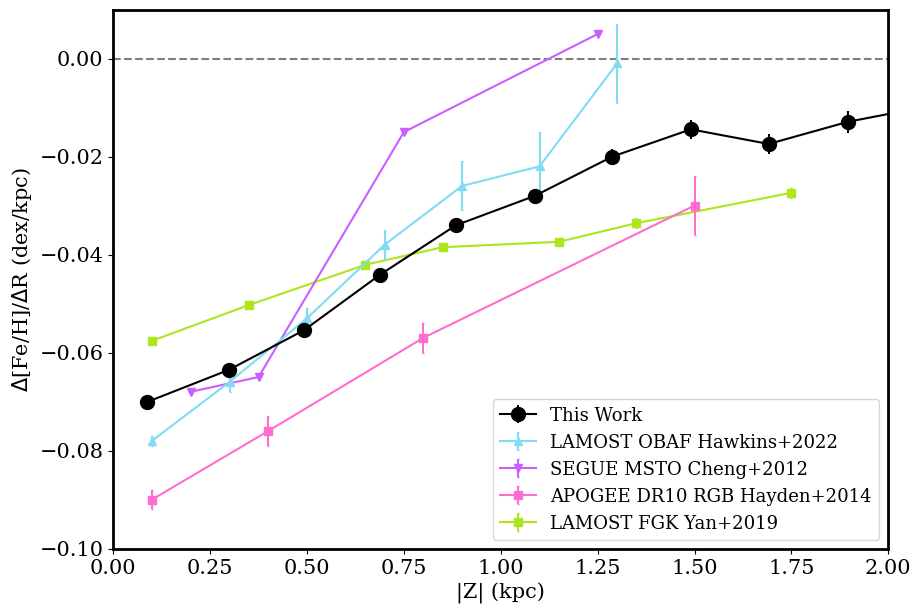}
	\caption{ The radial metallicity gradient as a function of absolute vertical height Z above (and below) the plane. The gradient derived in this work is represented by the black dots, compared to a variety of other studies that use different tracers [\citet{Cheng2012} (purple triangles), \citet{Hawkins2022} (blue triangles), \citet{Hayden2014} (pink squares), \citet{Yan2019} (green squares)]. Consistently, \Mgrad\ starts off at its most negative in the plane and shallows out with greater distances from the disk. A vertical line is plotted at \Mgrad=0 to illustrate where the radial gradient is no longer negative. The points derived in this work lie generally in the middle of the other studies conducted.}
	\label{fig:fehrvsz}
\end{figure}

The density of stars in the stellar disk of the Milky Way is thought to noticeably decline at about $\sim15.5$ kpc away from the Galactic center \citep{Momany2006,Reyle2009,Carraro2010}. The inner radii of the Galactic disk are thought to be dominated by the bar, typically believed to extend out to $\sim3.5$ kpc \citep{Hammersley1994,Wegg2015,Lucey2023}. Thus to minimize the chemical and dynamical effects driven by the bar in the most central region, and the drop in the density of stars in the furthest regions, we adopted a Galactocentric radius cut of $3.5 < R < 15.5$ kpc to our final thin disk sample. We place a 0.3 kpc cut on Z$_{\text{max}}$ to ensure that we are only selecting bona fide thin disk stars with orbits confined to the Galactic plane. Finally, we remove any stars in clusters as identified by \citet{Hunt2023} which reduced our sample by $\sim 200$ stars. After these final selection cuts, we have a planar thin disk sample of 25,404 stars with the median uncertainties for \Teff, \logg, and \feh\ being 33K, 0.07 dex, and 0.03 dex, respectively.


\section{Methods} \label{sec:Methods}

\begin{deluxetable}{cccc}
\tablecaption{Radial Metallicity Gradient Parameters for $\left| \text{Z} \right|$ bins in the full thin disk sample}
\label{tab: fehr table}
\tablehead{
\colhead{$\left| \text{Z} \right|$} & \colhead{$\Delta$[Fe/H]/$\Delta$R} & \colhead{$\sigma\Delta$[Fe/H]/$\Delta$R} & \colhead{N$_{\text{stars}}$} \\
\colhead{(kpc)} & \colhead{(dex/kpc)} & \colhead{(dex/kpc)} & \colhead{}
}
\startdata
0.1 & -0.0672 & 0.0004 & 40434 \\
0.3 & -0.0579 & 0.0005 & 42167 \\
0.5 & -0.0492 & 0.0006 & 35415 \\
0.7 & -0.0389 & 0.0007 & 25742 \\
0.9 & -0.0274 & 0.0009 & 14874 \\
1.1 & -0.0218 & 0.0011 & 9051 \\
1.3 & -0.0153 & 0.0013 & 5780 \\
1.5 & -0.0132 & 0.0015 & 3573 \\
1.7 & -0.0126 & 0.0016 & 2298 \\
1.9 & -0.0107 & 0.0018 & 1580 \\
\enddata
\end{deluxetable}

The behavior of the metallicity of stars throughout the Galactic disk can be best characterized by a negative linear gradient \citep[e.g.][]{Janes1979,Rolleston2000}. Using our planar thin disk sample, we derived the radial and vertical metallicity gradients stars as follows.

The gradients are initially represented as linear functions in which the gradients are represented by $m_{R}$ and $m_z$ in Equations \ref{eq:Mgrad} and \ref{eq:Vgrad}. In this, [Fe/H]$_{\text{R}}$ is the metallicity at a certain Galactocentric radius (R) and $m_{R}=$\Mgrad. To account for the metallicity at the Galaxy's center, we introduce the term $b_{R}$:

\begin{equation}
\mathrm{[Fe/H]_R }= m_{R} \mathrm{R}   + b_{R}
\label{eq:Mgrad}
\end{equation}

Similarly, we model the vertical metallicity gradient as:
\begin{equation}
\mathrm{[Fe/H]_Z}= m_{Z} \mathrm{|Z|}   + b_{Z}
\label{eq:Vgrad}
\end{equation}

where $m_z$ and $b_z$ are constants.

To find the gradients, we fit  linear models to the equations above with \texttt{scipy.stats.linregress} \citep{scipy2020} which calculates a linear least-squares regression for two sets of measurements (in our case Galactocentric radius/height and \feh). This returns the slope of the line (metallicity gradient), the standard error of the gradient, and the y-intercept (metallicity at the Galactic center/midplane). When computing the vertical gradient, we use our full thin disk sample without a radius or height cut. The resulting gradients are in Section \ref{sec:radial_grad}. In Tables \ref{tab: fehr table} (and \ref{tab: fehz table}), we compute the linear (vertical) metallicity gradients in radial (Galactic height) bins.

We use our 1D models to create 2D metallicity residual maps to search for potential signature of azimuthal variations.
We subtract the model abundances (in which the abundances are exactly equal to the linear gradient) from the observed \feh\ of each star in our sample. If the linear gradient is the only chemical feature of the disk, we would expect to see stochastic noise. If there is structure in this noise, then there are other processes driving the chemistry of the disk. We plot these residuals and investigate this azimuthal substructure in Section \ref{subsec:azvars}. When comparing the metallicity excess to azimuth, it is important to note that we define the line of sight from the Galactic center through the Sun to be an azimuth of $\phi(\pi) = 1$.

\begin{figure}	\includegraphics[width=1\columnwidth]{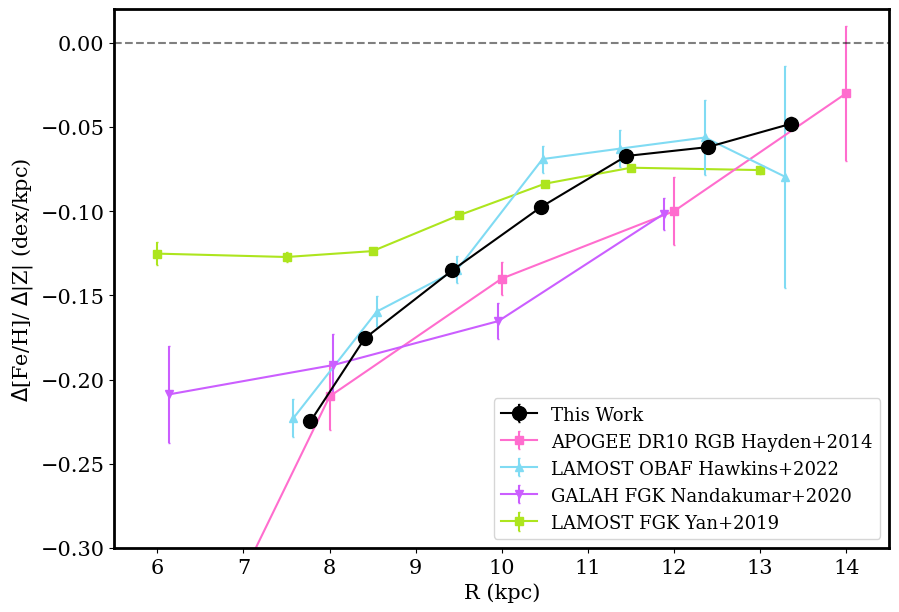}
	\caption{ The vertical metallicity gradient as a function of Galactocentric radius. The gradient derived in this work is depicted by black dots, whereas the other colored points represent the gradient as determined by different tracers in other studies [\citet{Hayden2014} (pink squares), \citet{Hawkins2022} (blue traingles), \citet{Nandakumar2020} (purple triangles), \citet{Yan2019} (green squares)]. The vertical metallicity gradient is at its most negative closest to the Galactic center and shallows out (\Vgrad\ approaches 0) as distance increases.}
	\label{fig:fehzvsr}
\end{figure}

The linear gradient is a useful approximation for \feh, however this does not necessarily hold true for all elements (see Appendix \ref{appendix:elemgrad}). To attempt to account for any non-linearity in our sample of elements, we employed a `running median' method. We began by calculating the median [X/Y] abundances in 0.2 kpc radial bins within the bounds of $\sim 6 - 15$ kpc. In these bins, we then took the median [X/Y] value and following the linear method, subtracted $Data - Median$ to quantify how the individual stars may deviate azimuthally from the median [X/Y] abundance. However, when this approach was taken, the radial bins did not sample azimuthal angles isotropically due to \apogee\ only probing one part of the disk. Thus, for the context of this work we only select elements that behave monotonically with respect to Galactocentric radius and move forward with the linear gradient method in Section \ref{subsec:otherelems}.


\section{Results and Discussion} 
\label{sec:result}

\begin{deluxetable}{cccc}
\tablecaption{Vertical Metallicity Gradient Parameters for R bins in the full thin disk sample}
\label{tab: fehz table}
\tablehead{
\colhead{$\left| \text{R} \right|$} & \colhead{$\Delta$[Fe/H]/$\Delta$Z} & \colhead{$\sigma\Delta$[Fe/H]/$\Delta$Z} & \colhead{N$_{\text{stars}}$} \\
\colhead{(kpc)} & \colhead{(dex/kpc)} & \colhead{(dex/kpc)} & \colhead{}
}
\startdata
7.5 & -0.2307 & 0.0026 & 29634 \\
8.5 & -0.1785 & 0.0022 & 49208 \\
9.5 & -0.1383 & 0.0025 & 33837 \\
10.5 & -0.0951 & 0.0026 & 22099 \\
11.5 & -0.0632 & 0.0028 & 14729 \\
12.5 & -0.0617 & 0.0030 & 8025 \\
13.4 & -0.0518 & 0.0034 & 3090 \\
\enddata
\end{deluxetable}

In this section we present our radial and vertical metallicity gradients (Section \ref{sec:radial_grad}), as well as azimuthal [Fe/H] variations throughout the thin disk (Section \ref{subsec:azvars}). We uncover azimuthal variations in elements beyond Fe in Section \ref{subsec:otherelems}. To investigate which mechanisms may be responsible for these variations, we bin our sample by age to explore the age dependence on azimuthal variations (Section \ref{subsec:ages}) and close out the section by examining the link between metallicity excess and several dynamical properties (Section \ref{subsec:kinematics}).



\subsection{Metallicity Gradients} \label{sec:radial_grad}

\begin{figure*}
\begin{center}
	 \includegraphics[width=2\columnwidth]{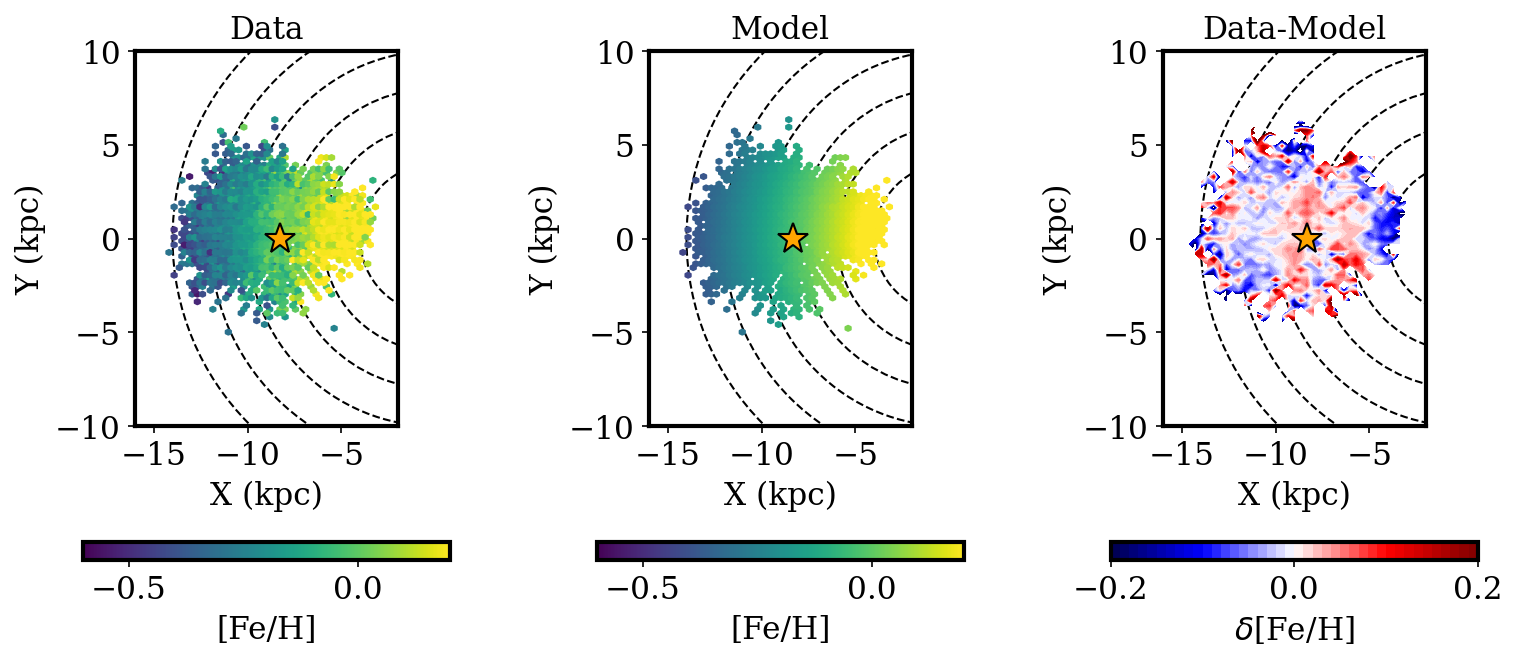}
	\caption{Left: The full \apogee\ planar thin disk sample colored by the metallicity. Middle: Each position of the data points colored by a model gradient of \Mgrad\ = \mygrad $+ 0.546$ dex/kpc. Right: the residuals of the observed \feh\ abundances and the linear model abundances. }
	\label{fig:fehgrad}
\end{center}
\end{figure*}

\begin{figure}
	 \includegraphics[width=0.9\columnwidth]{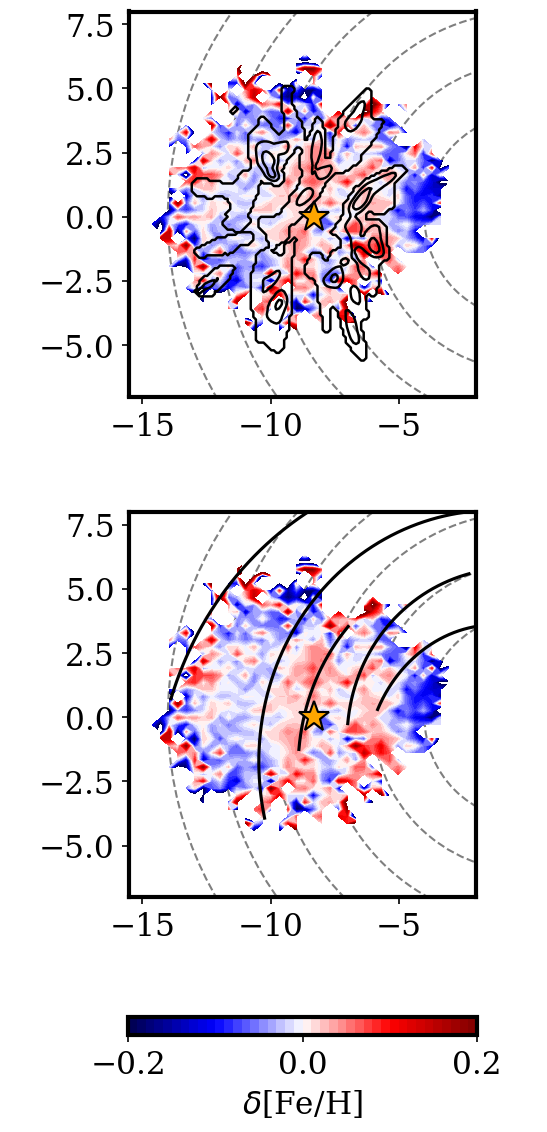}
	\caption{The azimuthal metallicity variations ($\delta$[Fe/H] in the last panel of Figure \ref{fig:fehgrad}) as compared to different determinations of the spiral arms. The black contours in the top panel represent the spiral arms as derived by \citet{Poggio2021} using main sequence stars in \gaia\ and the solid black lines in the bottom panel are the spiral arms as determined by  \citet{Reid2019} using high-mass star-forming regions. }
	\label{fig:spiralarms}
\end{figure}

Our first step in this analysis was to measure the radial (\Mgrad) and vertical (\Vgrad) metallicity gradients in our planar thin disk sample that have been previously observed with different tracer populations. Employing the methods outlined in Section \ref{sec:Methods}, we obtained a radial metallicity gradient with our planar thin disk sample of$\sim \mygrad\ \pm 0.0004$ dex/kpc (Figure \ref{fig:feh vs r}) as well as a vertical metallicity gradient of $\sim -0.164 \pm 0.001$ dex/kpc. Overall our gradients match within reason to other recently calculated gradients. \citet{Hawkins2022} derived a radial metallicity gradient of $\sim -0.078 \pm 0.001$ dex/kpc in the Galactic disk and a vertical metallicity gradient of $\sim - 0.15 \pm 0.01$ dex/kpc with an OBAF-type stellar sample from the LAMOST survey. Using the sixteenth data release of \apogee, \citet{Eilers2022} determined a radial gradient of $\sim -0.057 \pm 0.001$ dex/kpc. \citet{imig2023} calculated the running median radial metallicity gradients with \apogee\ DR17 ASPCAP parameters and found \Mgrad $=-0.064 \pm 0.001$ dex/kpc in their low-$\alpha$ sample at $|Z| \leq 0.25$, which is in good agreement with our gradient of $\sim \mygrad\ $ dex/kpc. We note one of the main distinctions in our sample compared to other works in the literature is the cut we imposed on Z$_{\text{max}}$ as opposed to present day height above the plane. 

\begin{figure*}
	 \includegraphics[width=2\columnwidth]{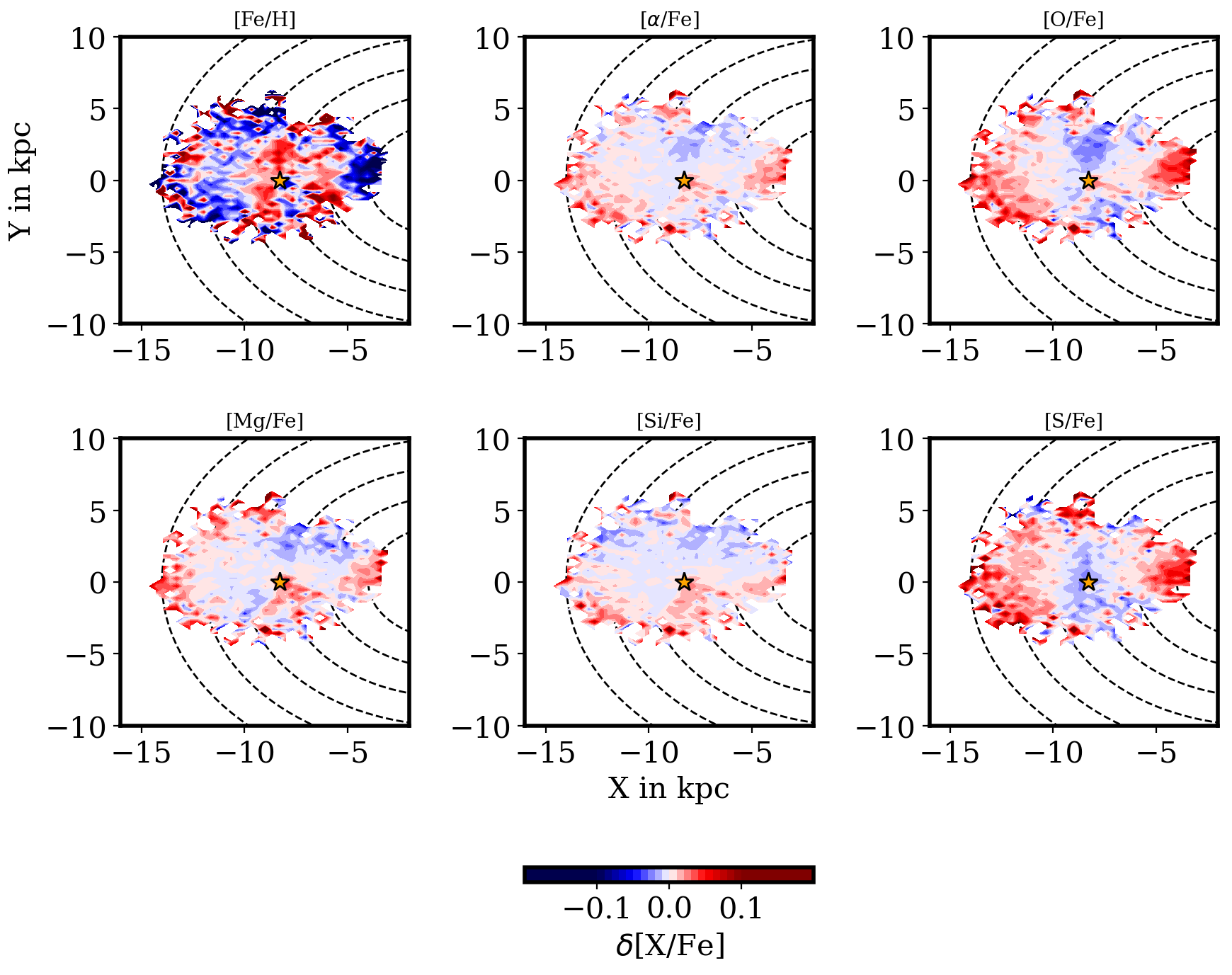}
	\caption{ The $Data-Model$ panel for the elements we deem to behave monotonically with respect to Galactocentric radius. The contours are colored by the shared [X/Fe] colorbar to illustrate the varying intensities of the deviations from the linear radial gradient. The second panel represents the average [$\alpha$/Fe] abundance, with the individual $\alpha$-elements in the following panels. The elements with the most saturated contours, such as \feh, showcase the most exaggerated deviation from the radial gradients. The $\alpha$-elements appear to be loosely anti-correlated with \feh, following predictions from the difference in timescales between events that mainly produce $\alpha$-elements (Type II SN) and events that mainly produce \feh\ (Type Ia SN). }
	\label{fig:allmeds}
\end{figure*}

The radial metallicity gradient is illustrated in Figure \ref{fig:feh vs r}. The metal enhancement of the inner Galaxy as compared to the outer Galaxy could point to the `inside-out' formation theory \citep{Larson1976,Kobayashi2011}. Starting at the inner Galaxy (i.e., R $\sim$3.5 kpc), the metallicity decreases linearly $-$ for every kiloparsec moving towards the outer disk, the global metallicity decreases by $\sim0.07$ dex.

Figure \ref{fig:fehrvsz} explores how \Mgrad\ varies with height above (and below) the Galactic plane. We separate the full thin disk sample into 0.2 kpc steps of |Z| and plot these results along with a select few other studies that use different tracer populations: LAMOST OBAF-type stars \citep{Hawkins2022}, SEGUE main sequence turn-off stars \citep{Cheng2012}, \apogee\ red giant stars \citep{Hayden2014}, and LAMOST FGK-type stars \citep{Yan2019}. Regardless of tracer population, all studies show a clear trend that as the distance from the Galactic plane increases, the radial metallicity gradient becomes more shallow, approaching zero. Physically, this means that the radial metallicity gradient is a prominent feature in the disk, but with increasing height from the Galactic plane the radial gradient is no longer the dominant observed relation and the vertical metallicity gradient starts to take over. We tabulate these results in Table \ref{tab: fehr table}. This work's \Mgrad\ vs. |Z| trend lies generally in the center of the other studies' trends. When compared with \citet{Hayden2014} which used the same survey as this work, merely an earlier data release (DR10 compared to our DR17), our work seems to be shifted up by $\sim0.01$ dex. This is not surprising due to the results of \citet{Jonsson2018} showing that \apogee\ data releases can have systematic differences of less than 0.05 dex.

Recent studies have shown that there is a vertical metallicity gradient that varies as a function of Galactocentric radius. \citet{Onal2016} found that the radial gradient is flat within 0.5 - 1 kpc of the plane and then becomes positive greater than 1 kpc away from the plane, suggesting that there is a vertical metallicity gradient in the Galaxy. To derive the vertical gradient, we applied the same methodology used for the radial metallicity gradient, obtaining \Vgrad\ $\sim -0.164 \pm 0.001$ dex/kpc. 

To quantify how the vertical gradient changes as a function of Galactocentric radius, we follow a similar methodology to the evaluation of the radial gradient varying with Galactic height, this time splitting our full thin disk sample of stars into radial bins of 1 kpc (Table \ref{tab: fehz table}). In Figure \ref{fig:fehzvsr}, we show our derived vertical metallicity gradient using the \apogee\ red giants, denoted by black circles. Other works that we compared to include vertical gradients obtained from \apogee\ red giant stars \citep{Hayden2014}, LAMOST OBAF-type stars \citep{Hawkins2022}, GALAH FGK-type stars \citep{Nandakumar2020}, and LAMOST FGK-type stars \citep{Yan2019}. We find that the vertical metallicity gradient is heavily correlated with Galactocentric radius in that \Vgrad\ approaches zero with increasing distance from the Galactic center.  


\subsection{Azimuthal Variations in \Mgrad} \label{subsec:azvars}

\begin{figure*}
\begin{center}
	 \includegraphics[width=2\columnwidth]{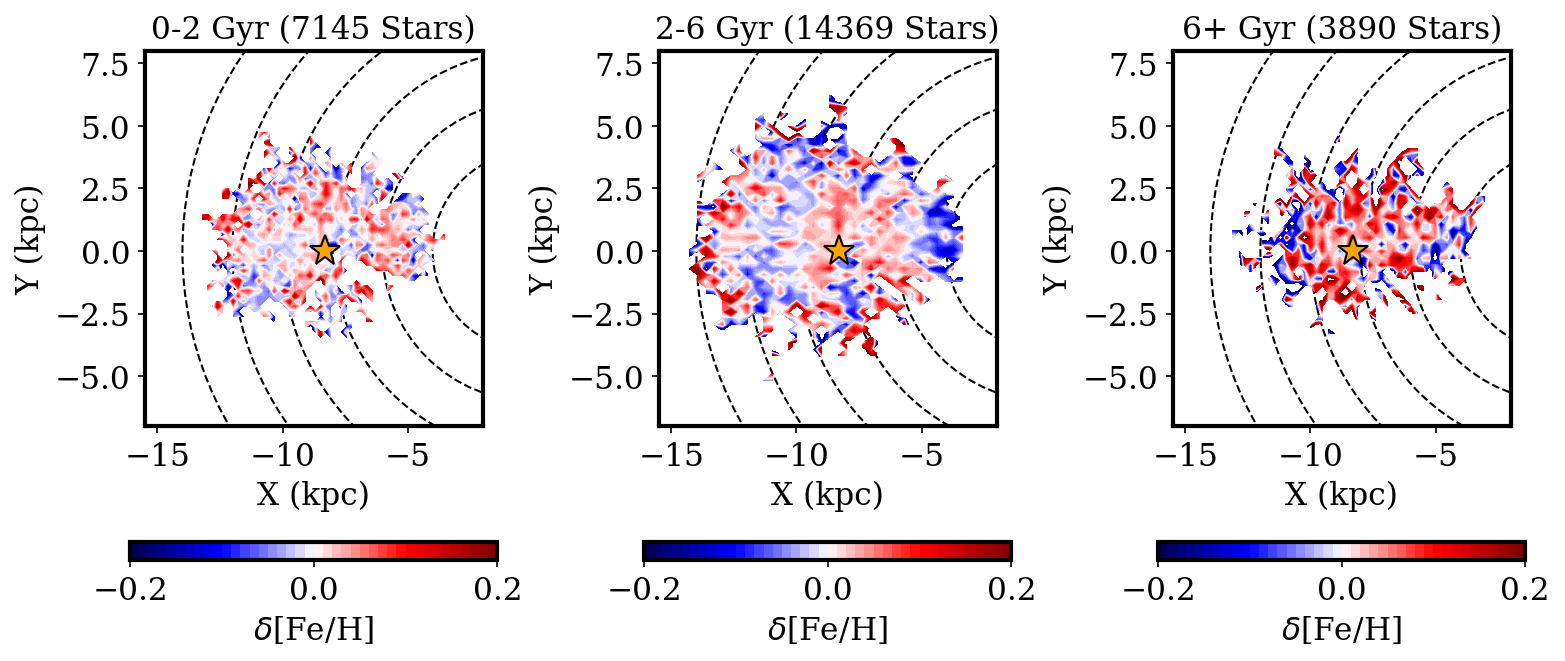}
	\caption{ This figure is similar to Panel 3 of Figure \ref{fig:fehgrad} increasing age groups of $0-2, 3-6,$ and $6+$ Gyr from left to right with 7297, 14419 and 3890 stars respectively. The color bars are constant throughout all panels, thus the youngest age group has the least amount of contrast and represents the smallest deviations from the linear gradient. Conversely, the oldest age group has the most saturated colors due to the larger variations from the modelled linear gradient.}
	\label{fig:ageresids}
\end{center}
\end{figure*}
In this section, we aim to characterize the angle-dependent deviations from the linear radial metallicity gradient in our planar thin disk sample and investigate any correlations with these deviations and the spiral arms. The discernible radial metallicity gradient has been recognized and characterized for decades \citep[e.g.,][]{Mayor1976,Andrievsky2002,Magrini2009,Boeche2013,Boeche2014,Cunha2016}. With increased sample size and precision provided by advances in instrumentation throughout the years, we are now able to peel back the gradient to see if there is any under-lying structure beneath this strong signature. 

To look beyond the metallicity gradient, we follow the steps outlined in Section \ref{sec:Methods} and compute the model linear \Mgrad\ gradient throughout the disk. We subtract this model off of the data to search for any structure in the residuals. When applied, we find variations in the \feh\ abundances that correlate with azimuthal angle.

We plot the top-down view of the Milky Way with our planar thin disk data in X and Y colored by \feh\ in Panel 1 of Figure \ref{fig:fehgrad}. We illustrate the model linear metallicity gradient in the second panel of Figure \ref{fig:fehgrad} and in the final panel we show the residuals that are found after subtracting the model from the data (labelled $\delta$[Fe/H]).

\begin{figure}
	 \includegraphics[width=1\columnwidth]{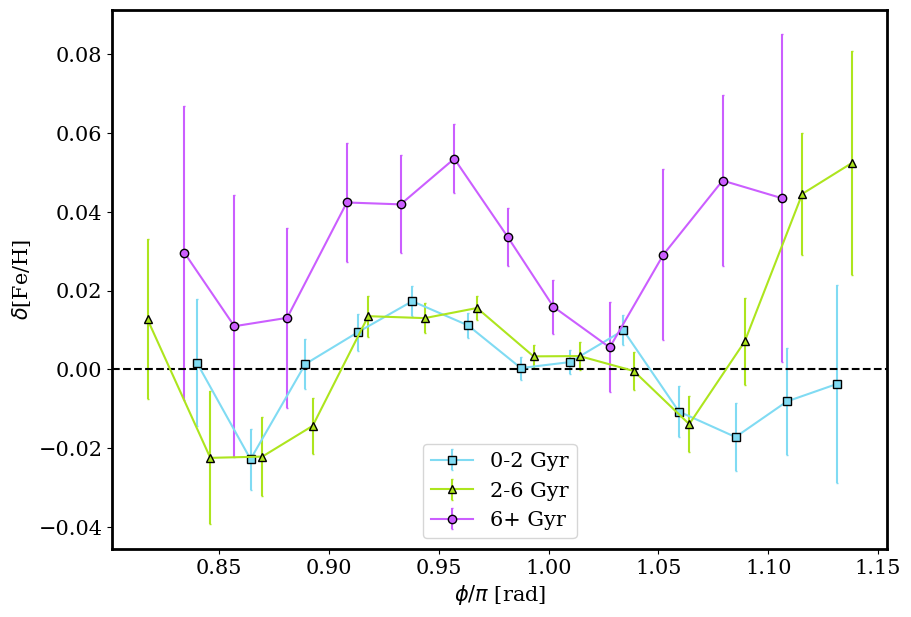}
	\caption{ This figure shows $\delta$[Fe/H] as a function of azimuthal angle ($\phi$). For reference, $\phi/\pi = 1$ is the line of sight from the Sun towards the Galactic center. The youngest age group, $0-2$ Gyr (blue squares), displays the smallest deviations from zero in $\delta$[Fe/H]. The intermediate age group, $2-6$ Gyr (green triangles), is the largest sample and has noticeable deviations from zero. The oldest age group, $6+$ Gyr (purple circles), varies the most dramatically in $\delta$[Fe/H] azimuthally. This depiction matches the theorized azimuthal variations, discussed further in Section \ref{subsec:ages}.}
	\label{fig:agegrads}
\end{figure}

There is a blue-red-blue-red pattern in $\delta$[Fe/H] shown in the final panel of Figure \ref{fig:fehgrad}. The red bins represent areas of the Milky Way that are more metal-rich than the model predicts and the blue bins are where the model over-estimates the stellar metallicities. These signatures remain the same when different cuts of Z$_{max}$ were tested. This oscillating pattern has been observed previously with different stellar surveys, such as \gaia\ DR3 stars in Figure 2 of \citet{Poggio2022}, with similar results to this work. \citet{Hawkins2022} use \lamost\ OBAF stars as well as \gaia\ DR3 stars, uncovering azimuthal variations in both of these populations, showing similar patterns in which the solar-neighborhood is more metal-rich (red) and the inner/outer galaxy are more metal-poor (blue). We explored whether the APOGEE selection function has an impact on recovered azimuthal variations using the selection function from the \texttt{gaiaunlimited} package \citep{gaiaunlimited}. We find that azimuthal variations are recoverable even with the impact of APOGEE's selection function.

To probe different formation pathways for this oscillating pattern, we first compare these results with the location of the spiral arms. \citet{Khoperskov2018} \citep[and][]{Grand2012} postulate that metal-rich stars from the center of the Galaxy may travel along the spiral arms, causing these filaments in the outer Galaxy to appear more metal rich. \citet{Khoperskov2018} show that azimuthal \feh\ variations in spiral galaxies would arise naturally if there is an initial negative radial metallicity gradient due to the migration of stars. To detect any correlation with the spiral arms, we plot our $\delta$[Fe/H] residuals and different determinations of the spiral arms in Figure \ref{fig:spiralarms}.

The black solid lines in the bottom panel of Figure \ref{fig:spiralarms} represent the spiral arms as determined by \citet{Reid2019} using the parallaxes and proper motions associated with high-mass star-forming regions using a Very Long Baseline Array, European VLBI Network, and the Japanese VLBI Exploration of Radio Astrometry project. They locate multiple arm segments with pitch angles ranging from 7\degree\ to 20\degree\ with the widths of the arms increasing with distance from the Galactic center. The black contours in the top panel of Figure \ref{fig:spiralarms} represent the spiral arm locations determined by \citet{Poggio2021} using the over-density of upper main sequence stars determined by \gaia\ DR3. These results were consistent with some of the \citet{Reid2019} arms, such as the Sagittarius-Carina spiral arm, while the geometry of arms with Galactic longitudes from 180\degree\ to 270\degree\ were significantly different from other spiral arm models. The metal-rich and metal-poor portions of the Milky Way in our work are not fully encompassed by either determination of the spiral arm locations, thus we look towards other processes that may have induced this azimuthal substructure.


\citet{Carr2022} quantified the effect that a Sagittarius (Sgr)-like dwarf galaxy would have on the kinematics and chemistry of the Milky Way-like disk upon first interaction. They did this by painting the particles in their N-body simulation of the disk with an negative radial [Fe/H] gradient. The passage of Sgr through the plane will cause radial rearrangement and disrupt stellar orbits. If the heating of orbits is due to a non-axisymmetric feature, such as Sgr, migration and mixing signatures will manifest as an approximate quadrupole in chemical azimuthal variations across the disk. In their present day snapshot of a simulated Milky Way being perturbed by a Sgr-like dwarf galaxy, \citet{Carr2022} found that the maximum departure from the \Mgrad\ gradient occurs in the solar annulus on the side of disk closest to the current position of Sgr, while the minimum is found adjacent to Sgr in the outermost annuli. These azimuthal variations in metallicity will be dependent on the model of Sgr chosen. It is interesting to note that in our sample the solar annulus appears to be more metal-rich, although we lack a complete view of the Galaxy to classify this definitively. 

While external influences like Sgr can cause azimuthal metallicity variations, studies have also shown that these features can arise through secular evolution with structures like the Galactic bar. \citet{filion23} simulated the effects of radial rearrangement in a barred galaxy and find substantial changes in the radii of stars that, when paired with the negative radial metallicity gradient, will induce azimuthal substructure similar to what we find with the \apogee\ stars \citep[see also][]{Dimatteo2013}. The radial rearrangement of stars due to the bar can be split into 3 zones to characterize the dynamics of the stars. The effects driven by the bar would be most prominent in the inner Galaxy where orbits evolve inward due to the angular momentum loss of the stellar orbits. Their intermediate zone is composed of stars moving both inwards and outwards producing no mean radial evolution. In the outer zone of their simulations, orbits generally evolve outwards (likely due to the net effect of the bar moving angular momentum outwards) and the trends are less aligned with the bar angle. Any migration of stars across the disk will produce azimuthal metallicity variations. In line with their findings, we see that the inner-disk stars of our sample have a lower metallicity than predicted by the gradient alone. 


To discern the responsible mechanisms for this substructure, it is crucial to confirm that this pattern is not unique to iron but extends to other elements as well. While the azimuthal [Fe/H] variations have been quantified before using \gaia\ \citep{Hawkins2022,Poggio2022}, \apogee\ gives us access to a suite of other elements. To confirm that this substructure is present in elements aside from iron, we evaluate other azimuthal chemical variations in the following section.

\subsection{Azimuthal Variations in Other Elements} \label{subsec:otherelems}

Here, we explore azimuthal variations in [Fe/H] and $\alpha$-elements (O, Mg, Si, S, Ca) in our planar thin disk sample. We choose to specifically highlight the $\alpha$-elements because the abundances of these elements in stars appear to behave monotonically with respect to Galactocentric radius, similar to [Fe/H]. For this methodology, we used our planar thin disk sample and plotted the median abundances of these elements in 0.2 kpc radial bins in Figure \ref{fig:allgrads}. As not all of these elements are monotonic with respect to Galactocentric radius, we focus on the $\alpha$-elements. We fit linear gradients to these elements and subtracted off the linear model (explained further in Section \ref{sec:Methods}) to look for structure in the residuals (similar to [Fe/H]), shown in Figure \ref{fig:allmeds}. 

In Figure \ref{fig:allmeds}, we show azimuthal variations with ranging intensities for all elements in our analysis, with variations detectable on at least the $\sim0.05$ dex-level. The second panel in this figure represents the average [$\alpha$/Fe] abundance, while the individual [$\alpha$/Fe] abundances are found in the following panels. We find that the $\alpha$-elements are loosely anticorrelated with \feh, which aligns with expected Galactic chemical evolution \citep[e.g.,][]{Tinsley1980}; in the disk where there is a $\delta$[Fe/H] excess, there is a $\delta$[$\alpha$/Fe] deficit. 

The anti-correlation between \feh\ and [$\alpha$/Fe] can be explained by the astrophysical processes (and timescales) that are largely responsible for the produciton of these elements. At early times, Type II SNe were able to effectively disperse $\alpha$-elements (e.g., O, Mg, Si, S, and Ca) as well as small amounts of Fe that were synthesized during the lifetimes of the first generations of stars. As the Galaxy continues to evolve, Type Ia SNe will then dominate, causing the [$\alpha$/Fe] abundances to lower due to more Fe-peak (Fe, Cr, Mn, Co, Ni) elements being made available to the next generation of stars \citep[e.g.,][]{Tinsley1980,Greggio1983,Woosley1992,Matteucci2001,Gonzalez2011}. Following this sequence of events, wherever there are [Fe/H]-excess regions, there will be [$\alpha$/Fe]-defecit regions and vice versa.


The presence of azimuthal variations in other elements confirms the existence of some process that is causing the observed chemical substructure throughout the Galactic disk. When looking for the responsible process, \citet{filion23} noted that focusing on younger populations ($\sim1-2$ Gyr) in the inner disk would help discern the role of secular evolution. Other simulations have used exclusively older stars to identify the cause of azimuthal chemical variations \citep{Dimatteo2013}. Consequently, we delve into the possible age-dependencies observed in azimuthal metallicity variations in the following section.


\subsection{Azimuthal Metallicity Variations by Age} \label{subsec:ages}

In this section, we aim to quantify the relationship between stellar age has and the magnitude of the azimuthal substructure found in Section \ref{subsec:azvars}. We are particularly interested in whether the amplitude of the azimuthal \feh\ variations are stronger in older or younger populations. We additionally aim to explore how the \feh\ gradient and azimuthal variations change as a function of age. For reference, it is important to note the difference between look-back time and present-day age. Look-back time is primarily used in simulations, where it  is the time elapsed from the `final'/present time back to a previous point in time, used to analyze historical states. In observations, we are limited to the present-day age of the object, or the total age of the object from when it formed to the present time. Thus, we aim to use the present-day ages of these stars and compare any trends in the ages with simulated works.

Using present-day age, the \citet{Debattista2024} simulations found that removing stars younger than 2 Gyr still produces azimuthal variations. This implies that azimuthal variations are not solely primordial in origin because older populations exhibit these variations as well. \citet{Bellardini2021} demonstrated that azimuthal scatter increases with increasing look-back time. This would indicate that stars that are born at earlier times will be born with stronger azimuthal variations due to the inhomogeneous interstellar medium (ISM) from which they form. While these two hypotheses appear to be contradictory, we cannot probe look-back times observationally so we cannot compare directly to \citet{Bellardini2021}, but we can compare our results to \citet{Debattista2024}. Consequently, in the rest of this section we make and test predictions about the observed relationship between azimuthal metallicity variations and stellar present-day age.

If the azimuthal variations are exclusively natal in origin (i.e., they are a result of variations in the gas-phase abundances at the time a population was born), then we would predict that younger populations would exhibit strong metallicity variations due to these populations forming more recently. To examine the effect that stellar age has on the strength of azimuthal metallicity variations, we applied the present-day ages from \apogee-\astronn\ to separate our sample by stellar age.


We separated the \apogee\ giants into 3 distinct age groups to explore the effects that age has on departures from the metallicity gradient. The age bins we selected are 0-2 Gyr, 2-6 Gyr, and $6+$ Gyr, with 7297, 14419 and 3890 stars, respectively. The strongest radial metallicity gradient appears in the middle age group with \Mgrad\ $\sim -0.0774 \pm 0.0005$ dex/kpc, outranking the youngest (\Mgrad\ $\sim  -0.0674 \pm 0.0006$ dex/kpc) and oldest (\Mgrad\ $\sim -0.0484 \pm 0.0012$ dex/kpc) age groups. We note that the radial metallicity trend is not well represented by a linear gradient in the oldest age bin.

The correlations we find with radial metallicity gradient and stellar age are consistent with the findings of \citet{Wang2019} who used the LAMOST survey to illustrate that stars of $4-6$ Gyr in age exhibit a steeper gradient than either younger or older stars. Similar work has been done with a CoRoT and \apogee\ sample by \citet{Anders2017} where their middle-aged population of stars had the steepest radial metallicity gradient, followed by the youngest then oldest population. This age-gradient trend is consistent with a systematic offset of $\sim -0.01$ dex/kpc between this work and \citet{Anders2017}.

After following the same methodology as Section \ref{subsec:azvars} and subtracting the data from the linear gradient model, Figure \ref{fig:ageresids} shows the $\delta$[Fe/H] residuals for each of the 3 age groups. Visually, the signatures seem to be the strongest in the panel containing the oldest stars, while the deviations lessen in intensity with decreasing age. Our methodology relies on the assumption that the radial metallicity gradient is linear in each age bin. Since the oldest population is not well-represented by a linear radial metallicity gradient, we do not draw any assumptions about the azimuthal structure for stars with ages of $6+$ Gyr. This population is included in Figures \ref{fig:ageresids} and \ref{fig:agegrads} for completeness.


In Figure \ref{fig:agegrads} we plot azimtuhal angle on the X axis and metallicity excess on the Y axis for each of the different age bins. We show that the most azimuthal variations are present in both young and intermediate age stars. There may be substructure in the oldest age bin, but these ages are not well-sampled azimuthally thus we refrain from drawing robust conclusions. For all age groups, the variations are minimized at sin$(\phi) = 1$, which is most likely due to the sample selection and the number of stars located along the line of sight from the Sun towards the center of the Galaxy. The future releases of SDSS V will provide better radial and azimuthal coverage to alleviate this problem.


We find that the strongest azimuthal \feh\ variations are apparent in solar-age stars. These results are in line with the simulated results of \citet{Debattista2024}. The authors found that azimuthal metallicity variations were still present when excluding younger populations, whereas we find the strongest azimuthal metallicity variations in our intermediate and older populations. Although azimuthal variations have been quantified in younger populations \citep[e.g., the OBA sample in][]{Hawkins2022}, the presence of the strongest azimuthal variations in our older populations suggest an important contribution from dynamical processes to the creation of azimuthal metallicity variations throughout the Galactic disk. Thus, in the following section, we delve into possible dynamical origins.


\subsection{Linking Chemistry to Dynamics} \label{subsec:kinematics}

\begin{figure*}
	 \includegraphics[width=2\columnwidth]{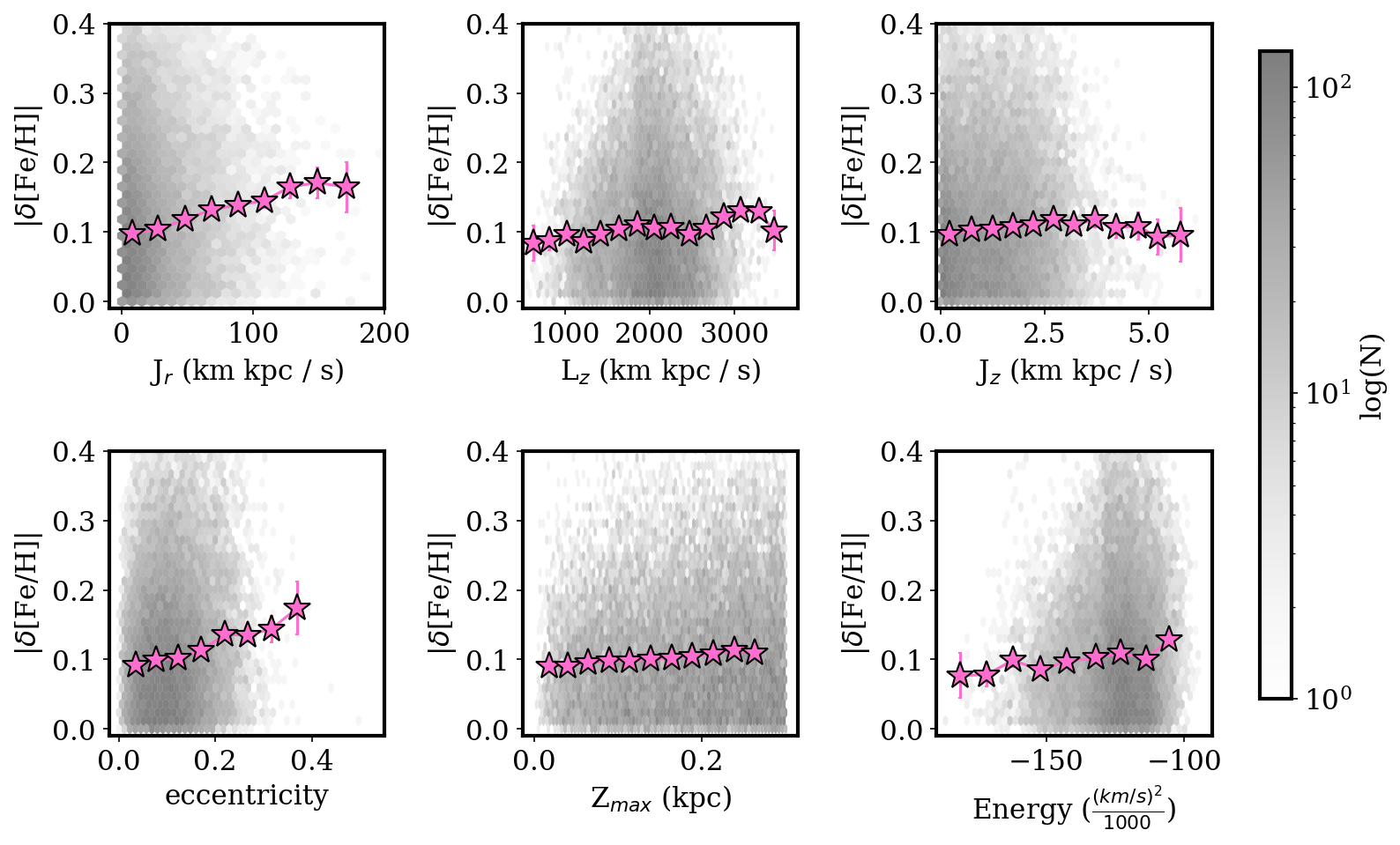}
	\caption{ The correlation between absolute metallicity excess ($\left| \delta \text{[Fe/H]} \right|$) on the y axes and different dynamical properties on the x axes. The hexagonal bins are colored by density with the pink stars representing the running medians in each panels with error bars showing the standard error on the median $|\delta$[Fe/H]$|$ in each eccentricity bin. In the J$_r$ and eccentricity panels, we see a clear increasing trend hinting that stars with high eccentricities and high radial actions contribute to the variations with the largest magnitudes.}
	\label{fig:dynamicalparams}
\end{figure*}

The driving mechanisms behind the chemical azimuthal substructure we see in the Galactic disk can be predominantly categorized into either radial migration/churning or kinematic heating/blurring. Radial migration \citep{sellwood2002} (also sometimes known as cold torquing), hereafter referred to as churning, is a dynamical process which altering stellar orbits through the interactions with non-axisymmetric features in the disk. More specifically, churning results in a change in the stellar orbits by changing the angular momenta of the orbits without adding any excess energy (e.g., stars on circular orbits experiencing churning will not see a change in eccentricity, only radii) \citep{Carr2022}. Blurring (also known as kinematic heating) heats an orbit without an increase in angular momenta of stellar orbits (i.e., there is an increase in eccentricity). Churning and blurring could be due to interactions with the spiral arms of the Galaxy \citep{jenkins1990}, interactions with the bar \citep{Schonrich2009,filion23}, and interactions with a satellite galaxy like Sgr \citep{Carr2022}. 

In this section, we aim to provide the initial basis of linking the azimuthal variations with Galactic dynamical properties to help elucidate the mechanisms responsible for these observed chemical signatures. While the goal of this paper is not to define the exact mechanisms responsible for azimuthal chemical variations, this initial linking to dynamics will motivate future studies. The dynamical properties of these stars could point to whether churning or blurring has a greater impact on the magnitude of the azimuthal variations. Thus we plot the determined radial action (J$_r$), angular momentum (L$_z$), vertical action (J$_z$), orbital eccentricity (e), maximum height above the Galactic plane (Z$_{\text{max}}$), and total energy to plot against absolute metallicity excess ($\left| \delta \text{[Fe/H]} \right|$). 

Figure \ref{fig:dynamicalparams} is a six panel plot showing the aforementioned parameters versus $\left| \delta \text{[Fe/H]} \right|$. The bins are colored by density with the pink stars representing the running medians in each panels with error bars showing the standard error in each eccentricity bin. There is a slight positive trend with $\left| \delta \text{[Fe/H]} \right|$ and J$_r$, suggesting that stars with a larger metallicity excess tend to have higher radial actions. This correlation with radial action follows the predictions of the \citet{Debattista2024} simulation\footnote{Note that those authors suggest that J$_r$ should be computed as a time-averaged quantity. We do not investigate the impact of time-averaging, and leave further investigation of this to future work.}. There are no obvious trends with the angular momentum, vertical action, maximum height, and total energy of stars when comparing to $\left| \delta \text{[Fe/H]} \right|$. To test if the older stars are solely responsible for the trends we are seeing in the dynamical properties, we redid this analysis for both our young and intermediate aged populations. These trends still hold in both age groups, thus we conclude that these dynamical trends are not just a byproduct of the oldest stars.

With the first and third panels of Figure \ref{fig:dynamicalparams}, we can make inferences about the relative importance of churning and blurring assuming that there is an initial negative metallicity gradient and no natal azimuthal variations. If churning is the primary mechanism causing azimuthal variations, we would predict that the largest $\left| \delta \text{[Fe/H]} \right|$ would be present at smaller eccentricities. This is because churning would largely cause stars to migrate from nearly circular orbits to nearly circular orbits of a different radius. If blurring is the primary mechanism causing azimuthal variations, we would expect the largest $\left| \delta \text{[Fe/H]} \right|$ values at larger eccentricities (and radial actions) because blurring is generally characterized by an increase in eccentricity. 

In the first and third panels of Figure \ref{fig:dynamicalparams}, we see larger $\left| \delta \text{[Fe/H]} \right|$ at higher eccentricities and radial actions. This could imply that blurring (heating of orbits) contributes a non-negligible amount to the mechanisms that are causing these observed azimuthal metallicity variations. In reality, multiple Galactic mechanisms are likely responsible for the azimuthal metallicity variations and more work is required to pin down mechanism responsible for the variations observed in this work and others \citep[e.g.,][]{Hawkins2022,Poggio2022,imig2023}.

While the goal of this paper is not to define the exact mechanisms responsible, this initial linking to dynamics will motivate future work on mock observations of simulations to directly compare different dynamical processes. We show that working in orbit-space, rather than present-day positions, can provide valuable insight into the dynamical properties that are driving the chemistry of the Galactic disk.

\section{Summary} \label{sec:summary}

Stellar spectroscopists have been using chemical cartography to map the chemistry of the Milky Way for the past decade \citep{Hayden2015}. With the chemistry of stars, we are able to discern what observable patterns are present throughout the Galactic disk and identify the key processes that contribute to Galactic evolution. Some of the most prominent patterns throughout the disk are the negative vertical and radial metallicity gradients \citep{Mayor1976,Andrievsky2002,Magrini2009,Luck2011,Bergemann2014,Xiang2015,Yan2019,Hawkins2022}. These gradients support  theories such as `inside-out' formation, which would cause the stars in the Galactic center to form first and ignite rapid star formation, with the outer Galaxy forming at later times. This will lead to the inner Galaxy having more metal-rich stars and the outer Galaxy to be populated by metal-poor stars. Hiding under the radial \Mgrad\ gradient are azimuthal variations on the scale of $\sim~0.1$ dex \citep{Poggio2022,Hawkins2022}. To constrain the properties of these azimuthal variations, we aim to confirm the \feh\ azimuthal variations in \apogee\ DR17, characterize how the variations interplay with stellar age, identify if azimuthal variations exist in elements other than \feh, and attempt to link the azimuthal variations to dynamical properties.

To accomplish these tasks, we model the linear radial metallicity gradient in the planar thin disk and subtract this off from the actual stellar metallicities to look for azimuthal substructure in \feh\ and $\alpha$-elements. We then separate our sample by age in to 3 subsets: 0-2 Gyr, 2-6 Gyr, and 6+ Gyr. We then quantified how metallicity excess ($\delta$[Fe/H]) changes with respect to radial action (J$_r$), angular momentum (L$_z$), vertical action (J$_z$), orbital eccentricity (e), maximum height above the Galactic plane (Z$_{\text{max}}$), and total energy. Based on these methods, we present the following results:

\

\begin{enumerate}
\item A radial metallicity gradient (\Mgrad) of $\sim \mygrad\ \pm 0.0004$ dex/kpc is found throughout the kinematic thin disk ( 3.5 $<$ R $<$ 15.5 kpc and Z$_{\text{max}} <$ 0.3 kpc) of the Milky Way. We additionally find a vertical metallicity gradient (\Vgrad) of $\sim -0.164 \pm 0.001$. Both gradients are in fairly good agreement with previous studies that use different tracer populations (Section \ref{sec:radial_grad}).

\item Azimuthal variations are found throughout the disk in \feh. These deviations were quantified by modelling the linear gradient and subtracting the model from the data to illustrate the \feh-poor to \feh-rich oscillating azimuthal pattern in the disk. These results are consistent with other azimuthal studies that have used different datasets or tracer populations \citep{Poggio2022,Hawkins2022}. While previous studies have found a correlation with azimuthal metallicity variations and the location of the spiral arms, we find some deviations to that in this work. We argue this is worth further investigation over a wider range of azimuthal angles (Hackshaw et al. in prep) (Section \ref{subsec:azvars}).

\item Azimuthal substructure is seen in elements beyond iron in [$\alpha$/Fe]. While [Fe/H] exhibits the strongest variations from the gradient, the $\alpha$-elements (O, Mg, Si, S) still show azimuthal variations on the $\sim0.1$ dex-level. These $\alpha$-abundances manifest as anti-correlations with [Fe/H] in some sections of the Galactic disk, following the predictions by the production mechanisms between $\alpha$ and Fe-peak elements (Section \ref{subsec:otherelems}).


\item Azimuthal substructure varies by stellar age. Azimuthal variations are quantified in both the young (0-2 Gyr) and intermediate (2-6 Gyr) populations. These variations in the young population suggests that the origins of this substructure could be either natal or dynamical, while the azimuthal variations that are present in the intermediate age bin suggests that these variations likely stem from dynamical processes rather than natal processes. The statistics are too low in the oldest age bin to draw robust conclusions. This age dependence can be coupled with mock observed simulations to identify the possible dynamical origins of these azimuthal variations (Section \ref{subsec:ages}).

\item There appear to be discernible trends with absolute metallicity excess ($\left| \delta \text{[Fe/H]} \right|$) and different dynamical properties. There is a positive trend between $\left| \delta \text{[Fe/H]} \right|$ and J$_r$, as well as $\left| \delta \text{[Fe/H]} \right|$ and eccentricity. This implies that stars with high eccentricities and high radial actions contribute to the azimuthal variations with the largest magnitudes, hinting that blurring is an important dynamical process in the production of azimuthal \feh\ variations. Further investigation of these links can help definitively characterize the cause of azimuthal metallicity variations (Section \ref{subsec:kinematics}).
\end{enumerate}

Azimuthal chemical variations are an apparent feature of the thin disk of our Galaxy. However, the cause of this substructure is unknown, with hints lying among the age dependence of these variations, azimuthal variations in elements beyond iron, and the dynamics of these stars. To answer this open question, we recommend deeper exploration of azimuthal variations with data like the impending release of SDSS-V, providing better radial and azimuthal coverage of the Galaxy. Including the orbit-space perspective rather than only the positions is a key step to understanding what drives these processes. Connecting the observable signatures in the Galactic disk with with mock simulated observations will bring us one step closer conclusively identifying what drives Galactic evolution.




\vspace{5mm}

\software{astropy \citep{astropy:2022,astropy:2013},  
          \texttt{astroNN} \citep{leung2019}, 
          \texttt{gala} \citep{gala_software},
          \texttt{PyAstronomy} \citep{pyastronomy},
          \texttt{scipy} \citep{scipy2020},
          \texttt{matplotlib.pyplot} \citep{matplotlib},
          \texttt{numpy} \citep{numpy}
          }


\section*{Acknowledgements}
ZH thanks Catherine Manea, Maddie Lucey, and Malia Kao for invaluable insight on this project.

KH \& ZH acknowledges support from the National Science Foundation
grant AST-2108736 and from the Wootton Center
for Astrophysical Plasma Properties funded under the United States
Department of Energy collaborative agreement DE-NA0003843.
This work was performed in part at Aspen Center for Physics,
which is supported by National Science Foundation grant PHY1607611. This work was performed in part at the Simons Foundation
Flatiron Institute’s Center for Computational Astrophysics during
KH’s tenure as an IDEA Fellow.
This project was developed in part at the Gaia Fete, hosted by
the Flatiron Institute Center for Computational Astrophysics in 2022
June. This project was developed in part at the Gaia Hike, a workshop
hosted by the University of British Columbia and the Canadian
Institute for Theoretical Astrophysics in 2022 June. This project was developed in part at the 2023 Gaia XPloration, hosted by the Institute of Astronomy, Cambridge University.
CL acknowledges funding from the European Research Council (ERC) under the European Union’s Horizon 2020 research and innovation programme (grant agreement No. 852839).

Funding for the Sloan Digital Sky Survey V has been provided by the Alfred P. Sloan Foundation, the Heising-Simons Foundation, the National Science Foundation, and the Participating Institutions. SDSS acknowledges support and resources from the Center for High-Performance Computing at the University of Utah. The SDSS web site is \url{www.sdss.org}.

SDSS is managed by the Astrophysical Research Consortium for the Participating Institutions of the SDSS Collaboration, including the Carnegie Institution for Science, Chilean National Time Allocation Committee (CNTAC) ratified researchers, the Gotham Participation Group, Harvard University, Heidelberg University, The Johns Hopkins University, L’Ecole polytechnique f\'ed\'erale de Lausanne (EPFL), Leibniz-Institut f\"ur Astrophysik Potsdam (AIP), Max-Planck-Institut f\"ur Astronomie (MPIA Heidelberg), Max-Planck-Institut f\"ur Extraterrestrische Physik (MPE), Nanjing University, National Astronomical Observatories of China (NAOC), New Mexico State University, The Ohio State University, Pennsylvania State University, Smithsonian Astrophysical Observatory, Space Telescope Science Institute (STScI), the Stellar Astrophysics Participation Group, Universidad Nacional Aut\'{o}noma de M\'exico, University of Arizona, University of Colorado Boulder, University of Illinois at Urbana-Champaign, University of Toronto, University of Utah, University of Virginia, Yale University, and Yunnan University.

\appendix

\section{Thin Disk Sample Determination} \label{appendix:thindisk}
The thick-to-thin disk probability ratios (TD/D) were found for each star using the assumption that the space velocities U, V, and W have Gaussian distributions \citep{Bensby2014}, 

\begin{equation} \label{probs}
f = k \cdot \text{exp} \Big( -\frac{(U_{\text{LSR}} - U_{\text{asym}} )^2}{2\sigma_{\text{U}}^2} -\frac{(V_{\text{LSR}} - V_{\text{asym}} )^2}{2\sigma_{\text{V}}^2} -\frac{W_{\text{LSR}}^2}{2\sigma_{\text{W}}^2} \Big)   
\end{equation}

where the normalization factor $k$ is given by

\begin{equation} \label{probs}
k = \frac{1}{(2\pi)^{3/2}\sigma_{\text{U}} \sigma_{\text{V}}\sigma_{\text{W}}}   
\end{equation}

The characteristic velocity dispersions are represented by $\sigma_{\text{U}}, \sigma_{\text{V}}$ and $\sigma_{\text{W}}$ and $U_{\text{asym}}$ and $V_{\text{asym}}$ are the asymmetric drifts. Finally the thick (TD) to thin disk (D) ratio is found with

\begin{equation} \label{probs}
\text{TD/D} = \frac{X_{\text{TD}}}{X_{\text{D}}} \cdot \frac{f_{\text{TD}}}{f_{\text{D}}}
\end{equation}

\begin{figure*}
\begin{center}
\includegraphics[width=0.9\columnwidth]{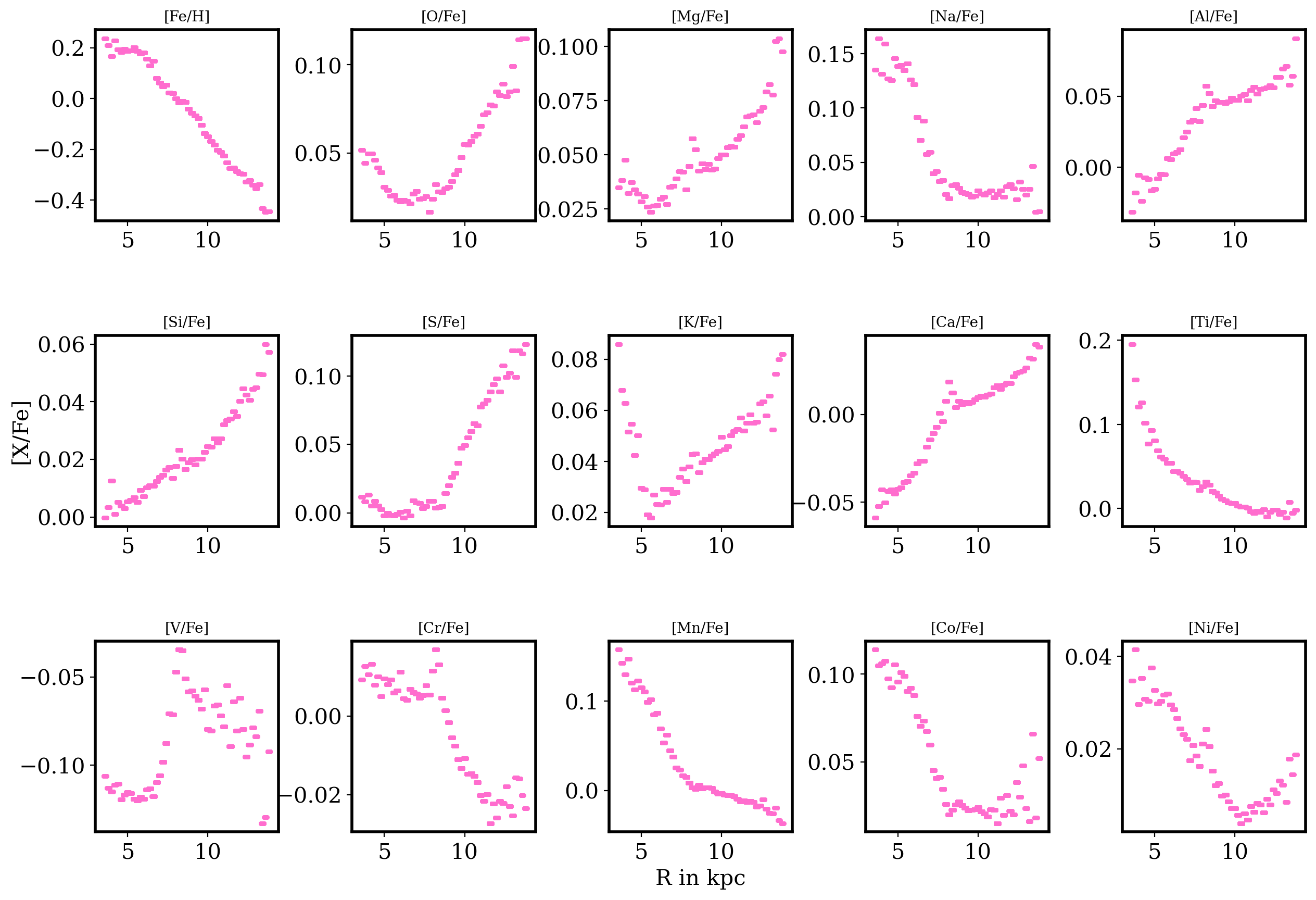}
\caption{The running median [X/Fe] abundances in 0.2 kpc radial bins for 15 elements in our sample. The high-fidelity elements (\feh, [O/Fe], and [Mg/Fe]) are featured on the top row. A number of these elements would not be best represented by a linear gradient, so we leave a majority of them out of our analysis. The elements included when searching for azimuthal variations in the linear gradient are the approximately linear elements: Fe, O, Mg, Si, S. To look at the $\alpha$-element behavior more holistically, we averaged the $\alpha$-element abundances in Figure \ref{fig:allmeds}.}
\label{fig:allgrads}
\end{center}
\end{figure*}

We characterize thin disk stars as any star with a thin disk probability $\geq 80 \%$.
\section{Elemental Gradients} \label{appendix:elemgrad}
This appendix contains the [X/Fe] median abundance trends in 0.2 kpc bins in Figure \ref{fig:allgrads}. Note the y-axis of the panels vary, but the elemental families show general trends as expected, such as the Fe-peak (Cr, Mn, Fe, Co, and Ni) exhibiting negative gradients. The $\alpha$-elements (O, Mg, Si, S, Ca) all have positive trends. These gradients match well with shape and scale to other studies that have looked at abundance trends throughout the Galactic disk \citep[for example, Figure 7 in][]{Eilers2022}.

Due to the fact that azimuths are not sampled isotropically in the 0.2 kpc radial bins, the methodology of calculating $\Delta$\feh\ from the abundance medians is inaccurate. Thus, we only choose elements with roughly linear and monotonic shapes to complete our analysis (Fe, O, Mg, Si, S).

\bibliography{bibliography}{}
\bibliographystyle{aasjournal}

\end{document}